# Transport Protocols in Cognitive Radio Networks: A Survey

**Xiaoxiong Zhong, Yang Qin\* and Li Li**

Key Laboratory of Network Oriented Intelligent Computation, Shenzhen Graduate School,
Harbin Institute of Technology, Shenzhen, 518055, P.R. China
[e-mail: {xixzhong, yqinsg} @gmail.com, lili8503@aliyun.com]
\*Corresponding author: Yang Qin

---

## *Abstract*

Cognitive radio networks (CRNs) have emerged as a promising solution to enhance spectrum utilization by using unused or less used spectrum in radio environments. The basic idea of CRNs is to allow secondary users (SUs) access to licensed spectrum, under the condition that the interference perceived by the primary users (PUs) is minimal. In CRNs, the channel availability is uncertainty due to the existence of PUs, resulting in intermittent communication. Transmission control protocol (TCP) performance may significantly degrade in such conditions. To address the challenges, some transport protocols have been proposed for reliable transmission in CRNs. In this paper we survey the state-of-the-art transport protocols for CRNs. We firstly highlight the unique aspects of CRNs, and describe the challenges of transport protocols in terms of PU behavior, spectrum sensing, spectrum changing and TCP mechanism itself over CRNs. Then, we provide a summary and comparison of existing transport protocols for CRNs. Finally, we discuss several open issues and research challenges. To the best of our knowledge, our work is the first survey on transport protocols for CRNs.

***Keywords:*** Transport protocol, cognitive radio networks, survey, spectrum efficiency

---

# 1. Introduction

**W**ith the demands of wireless technologies and applications, more and more spectrum resources are needed. With the current spectrum allocation policy, all of the spectrum bands are exclusively allocated for licensed users (i.e., primary users PUs), and violation from unlicensed users (i.e., secondary users SUs) is not allowed. This is the main factor that leads to spectrum under-utilization. The Federal Communications Commission (FCC) has indicated that temporal and geographical variations in the utilization of the assigned spectrum range from 15% to 85% [1]. Dynamic spectrum access (DSA) [2] is proposed to solve the critical problem of spectrum scarcity. This new research area foresees the development of CRNs.

The cognitive radio (CR) [3] principle has introduced the idea to exploit spectrum holes (i.e., bands) which result from the proven underutilization of the electromagnetic spectrum by modern wireless communication and broadcasting technologies. The exploitation of these holes can be accomplished by the notion of CRNs. CRNs have emerged as a prominent solution to improve the efficiency of spectrum usage and network capacity. There are three main cognitive radio network paradigms: underlay, overlay, and interweave [4]. The underlay paradigm allows cognitive users to operate if the interference caused to non-cognitive users is below a given threshold. In overlay systems, the cognitive radios use sophisticated signal processing and coding to maintain or improve the communication of non-cognitive radios while also obtaining some additional bandwidth for their own communication. In the interweave systems, the cognitive radios opportunistically exploit spectral holes to communicate without disrupting other transmissions. The interweave model is the one predominantly studied in the literature, and the de facto standard for CRNs. Note that some researchers classify the transmission modes for CRNs into two paradigms: overlay and underlay mode [5, 6]. To the best of our knowledge, majority of the research efforts in transport layer protocols for CRNs focus on interweave/overlay model, i.e., in CRNs, the SUs can opportunistically exploit frequency bands when the PUs do not occupy the bands. When a PU arrives, the SU must vacate the channel to the PU and continue to use another available channel, if any. This is due to the basic principle that the PUs have the exclusive priority to occupy the channel. Thus, we adopt the categories in [5, 6] throughout this paper.

Most of research works that have been conducted in CRNs concentrate on the two lower layers, tackling PHYsical (PHY) layer and/or media access control (MAC) layer issues, including the definition of effective spectrum sensing, spectrum decision and spectrum sharing techniques. However, the transport layer protocols for CRNs are in the nascent stages of development. To the best of our knowledge, there is no survey on transport protocols for CRNs in the current literature, and there is a need of such a survey to know the state-of-the-art.

Transmission control protocol (TCP) [7] is a dominant transport layer connection oriented and reliable end-to-end protocol. It provides reliable data delivery with the help of its flow control and congestion control mechanism. In order to improve TCP performance many modifications were proposed. However, some of the unique features of CRNs make TCP performance degrade. From the literature review reported in this paper, some of these unique features of CRNs are PU arrivals, spectrum sensing, spectrum changing, and heterogeneous available channels in SUs. It is well-known that TCP has a degraded throughput under wireless systems especially with a high packet loss rate because TCP assumes all packet loss is due to congestion and triggers rate reduction whenever packet loss is detected. In CRNs, there is a new loss-service interruption loss [5], due to the existence of PUs. These features will cause packet losses and timeouts which are mistakenly categorized as congestion losses by TCP. Recently, some transport protocols, which based the unique features of CRNs on evaluating or improving TCP performance, have been presented.

Our goal in this paper is to provide a clear overview of different proposals suggested by the research community for performance improvement of TCP in CRNs and provide a guide as to what are the possible directions for further improvements in this area. The main contributions of our work are listed as follows:

- We summarize the TCP's challenges in CRNs.
- We classify the current state of transport protocols into different categories and analyze and discuss them in a systematic way. In this way, it aims to provide an overview of the current state of TCP on CRNs. In the process, a classification of the proposals is provided to give the reader a new angle from which to view the work on TCP in CRNs.
- We outline several major open issues and research challenges, which mainly consider the features of CRNs.

This paper is organized as follows. Section 2 gives an overview of TCP working mechanism. Section 3 outlines the challenges TCP faces in CRNs. Section 4 presents a survey of the approaches available to evaluate/improve the performance of TCP in CRNs and provides a taxonomy of these approaches. Section 5 provides suggestions for possible directions for future study seeking to improve the performance of TCP in CRNs. Section 6 concludes our work.

## 2. An Overview of TCP Working Mechanism

The transport layer is the key to understanding layered protocols. It provides various services, the most important of which is an end-to-end, reliable, connection-oriented byte stream from sender to receiver. TCP is a window-based reliable transport layer protocol that achieves its reliability through sequence numbers and acknowledgements (ACKs). The sending and receiving TCP entities exchange data in the form of segments.

To ensure the reliable delivery of data segments, when the receiver receives a segment, it replies to the sender to acknowledge that the segment has been received correctly and to send the sequence number of next expected segment. More than one ACK identifying the same segment to be retransmitted is called a duplicate ACK. In the case that an out-of-order segment arrives at the receiver, a duplicate ACK is generated and sent back to the sender. After three duplicate ACKs, the sender assumes that the segment has been lost and retransmits it. Moreover, TCP also uses timeout to detect losses. After transmitting a segment, TCP starts a time down counter to monitor timeout occurrence. If timeout occurs before receiving the ACK, then the sender assumes that the segment has been lost. The lost segment is then retransmitted, and TCP initiates the slow start algorithm. The timeout interval is called retransmission timeout (RTO) and is computed according to [8]. By retransmitting the lost segment, TCP achieves reliable data transmission.

TCP uses an additive-increase multiplicative-decrease (AIMD) strategy for changing its window as a function of network conditions [7, 9]. As an illustration of how the congestion window works, as shown in Fig. 1. Starting from one packet, or a larger value, the window is increased exponentially by one segment for every non-duplicate ACK until the source estimate of network capacity is reached. This is the Slow Start (SS) phase, and the capacity estimate is called the SS threshold (*sthresh*). Once *sthresh* is reached, the Congestion Avoidance (CA) starts and it is increased linearly, i.e., it is increased by one segment for each Round-Trip Time (RTT). The window increase is interrupted when a loss is detected.

Two mechanisms are available for the detection of losses described above: the expiration of an RTO or the receipt of three duplicate ACKs. The source supposes that the network is in congestion and sets its estimate of the capacity to half the current window. Upon an RTO, *ssthresh* is set to the half of the current congestion window and the congestion window is reduced to 1 MSS (maximum segment size). The SS starts again, as shown in **Fig. 1**. Note that the sender reacts to three duplicate ACKs in a different way, which includes Fast Retransmit (FRXT) and Fast Recovery (FRCV). Tahoe **[7]**, the first version of TCP to implement congestion control, at this point sets the window to one packet and uses SS to reach the new *ssthresh*. When the loss is detected via timeout, SS is unavoidable since the ACK clock has stopped and SS is required to smoothly fill the network. However, in the FRXT case, ACKs still arrive at the source and losses can be recovered without SS. This is the objective of the TCP versions (Reno **[10]**, New Reno **[11]**, SACK **[12]**, Vegas **[13]**, etc.) that call a FRCV algorithm to retransmit the losses while maintaining enough packets in the network to preserve the ACK clock. Once losses are recovered, this algorithm ends and normal CA is called. If FRCV fails, the ACK stream stops, a timeout occurs, and the source resorts to SS as with Tahoe.

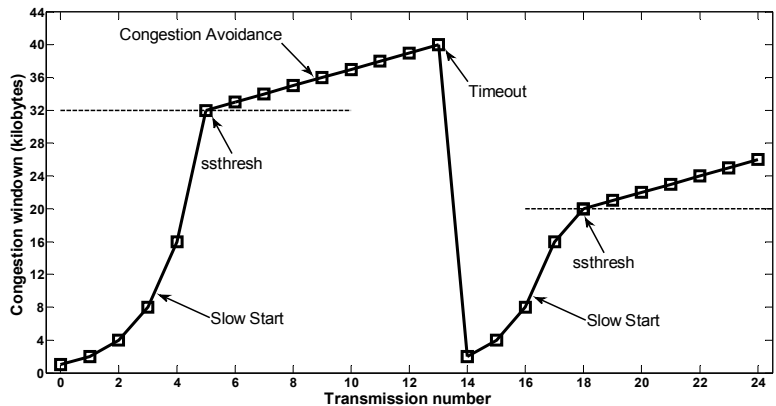

**Fig. 1.** TCP congestion window dynamics **[14]**.

## 3. TCP's Challenges in CRNs

It is a known fact that TCP is inherently inefficient on wireless networks. In a wireless environment, packets may be lost due to congestion or errors of channel. The performance of TCP worsens in CRNs, because TCP has to face new challenges due to several reasons specific to these networks: (i) PU behavior, (ii) spectrum sensing, (iii) spectrum changing, and (iv) TCP itself.

Here, we describe the models used in many existing works for TCP performance analysis in CRNs, which adopt an interweave model in CRNs, i.e., the nodes in the CRNs can only transmit when the PUs are not active. When PU appears on a channel, the SU should vacate it for the PU. The SU's communication mainly depends on the PU's activities, which is the biggest difference between CRNs and traditional wireless networks such as Ad hoc networks and Mesh networks. The process of sharing spectrum between PUs and SUs is shown in **Fig. 2**.

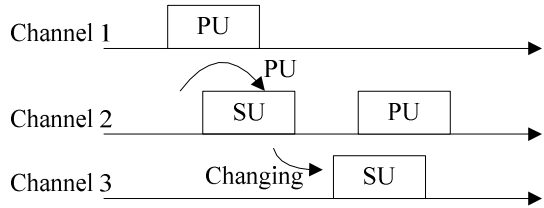

**Fig. 2.** Spectrum state in CRNs.

### 3.1 PU Behavior

In CRNs, the PUs have the highest priority to use the spectrum, whether SUs are in the spectrum sensing and data transmitting phase. On detecting the presence of PUs, the SUs must immediately cease their operations on that channel and search for another vacant channel, which would cause a large amount of packet loss (interruption loss). It will cause a congestion and packet loss, and then leads to significant TCP performance degradation. If there are no vacant channels currently, SUs have to wait until the next sensing cycle comes. In this case, SUs will have to disconnect the TCP connections. During the waiting period, retransmission timer will be triggered. If an idle channel is available through sensing operations, the SUs can continue to transfer data packets.

### 3.2 Spectrum Sensing

In CRNs, the spectrum sensing is a periodical process that monitors the current channel over a pre-defined sensing duration for occurrence of a PU. From SUs view, the channel state includes two states: sensing state and transmitting state. In sensing state, the link is in a virtual connection state; the nodes that perform spectrum sensing cannot transmit/receive data packets. If the sensing duration is long, it may result in a decrease in the number of transmitted packets. Also, PU detection errors may severely affect the transport protocol design.

### 3.3 Spectrum Changing

When a SU is in a transmission state on a channel, at this moment, if a PU arrives, it immediately occupies the channel for its transmission. Thus, the SU ceases its transmission and searches for the idle channels. If it currently has an idle channel in its channel list, it changes from the former channel to the idle channel. In this case, there may have a large variation in bandwidth, which will largely affect TCP performance. In spectrum changing state, it may induce interruption, which may increase the RTT or trigger a TCP's RTO.

At spectrum sensing/change intervals, the path becomes virtually disconnected, with the sensing or changing node unable to forward packets in either direction. This event may, inadvertently, trigger the TCP timeout conditions at the source thereby causing it to mistake the spectrum-related function as network congestion.

### 3.4 TCP Itself

In CRNs, the network topology frequently changes, due to spectrum sensing, spectrum decision and spectrum changing. Thus, the CRNs are extremely unstable. In the traditional TCP protocol, if a packet loss event occurs, TCP will not hesitate to think that this is network congestion. Congestion control technique must be carried out to avoid congestion collapse, reducing the congestion window and slowing down sending rate. There is a new type of packet loss in CRNs, called interruption loss, which is caused by PU's activity. If we perform the same operations on TCP (TCP reduces the congestion window to 1 segment and resets to the slow-start state) as traditional packet loss occurs, the TCP performance will degrade. This process frequently occurs due to frequent PU arrivals on the current channel. Thus, the traditional TCP mainly keeps a small congestion window in CRNs, the TCP does not provide very good performance for CRNs. As a result, if we can distinguish the interruption loss causing by PU arrivals, the TCP throughput will improve.

## 4. The Existing Transport Protocols for CRNs

### 4.1 Taxonomy of Available Proposals

TCP in CRNs has similarities with TCP in traditional wireless networks, but with the additional challenge of having to deal with the dynamic behavior of the PUs, and their effects on changing spectrum opportunities (SOPs) of SUs. Some efforts have been taken to improve the TCP performance in CRNs through optimizing the lower-layer parameters or modifying TCP itself.

We classify the existing transport protocols into two categories based on transmission environments (single hop or multi-hop), and in each category, the existing proposals are described and analyzed from two approaches, layered and cross-layer. Readers can refer to **Fig. 3** for an overview of the surveyed schemes. Note that, the DSASync [29] (classical for single hop), TP-CRAHN [34, 35] (classical for multi-hop) and TCPJGNC [48] (using network coding to mask losses) are described in more detail due to their specificities.

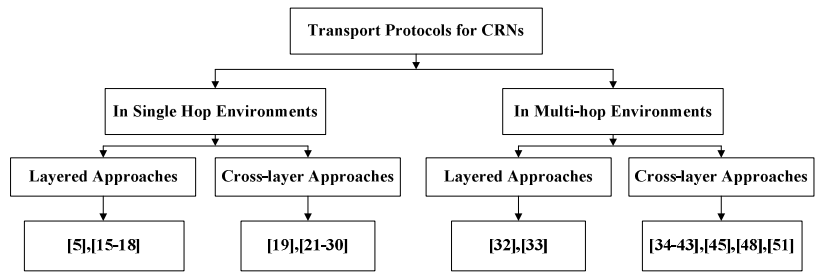

**Fig. 3**. Existing transport protocols for CRNs.

### 4.2 Transport Protocols for Single Hop Environments

In this section, we focus on TCP performance in single hop CRNs, which typically include wireless LAN and wireless cellular networks, from layered and cross-layer perspectives. The typical scenario is shown in **Fig. 4**, in which the sender is connected to a Base Station (BS) by means of a wired line, maybe going through the Internet, and the receiver is connected to the BS via a DSA link of varying capacity. It is a 1-hop topology between the CR nodes and the BS. A TCP connection is established between the sender and receiver, and the flow of TCP segments travels from sender to receiver, while TCP ACKs flow in the opposite direction. In this scenario, when PUs arrive, the transmission between BS and SUs is interrupted, the SUs will find an idle channel or wait for the operating channel to be idle state to transmit data packets. If the duration time is long, it will result in TCP performance degradation.

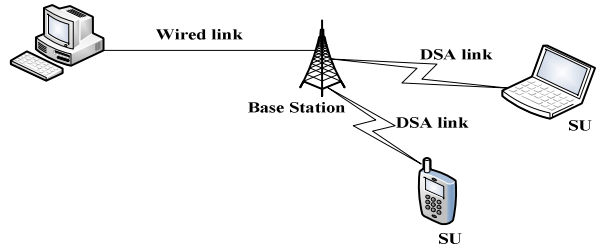

**Fig. 4.** The single hop network scenario.

## A. Layered approaches

Slingerland *et al.* [15] evaluated TCP (SACK, NewReno, Vegas) performance in DSA scenarios. They analyzed the effect of varying link capacity and PU detection errors on throughput of TCP in a DSA environment. Also, they concluded that the both NewReno and Vegas perform very well for all link models in the following situation: a large buffer is available in the BS and the receiver employs selective acknowledgements. In practice, the capacity of the BS buffer is limited, and thus BS overflow may occur, however, the authors did not consider it. In addition, they found that selective acknowledgments are essential to good performance in scenarios with large end-to-end delay.

Kondareddy *et al.* [16] modified the analytical model proposed in [15] to incorporate the delay caused by PU's and SU's traffic and PU detection errors. The PU and SU traffic are modeled using continuous-time Markov chains and the blocking probability of the SUs is calculated. Through extensive simulations, it shows that the proposed analytical model is efficient in capturing the dynamic nature of DSA networks.

In both cases, the authors provided insight on the effect of DSA links over conventional TCP performance, by using analytical and simulation tools. They found that the TCP performance is significantly degraded because of frequent interruptions due to the PU arrivals on the current channel. However, they did not provide a scheme for TCP performance improvement in CRNs.

Issariyakul *et al.* [5] evaluated the performance of TCP NewReno on CRNs. They only considered a new type of loss called service interruption loss, due to the existence of PUs. To capture the amount of PU activity, which determines the amount of service interruptions loss for SUs in CRNs, they defined load factor (*A*) as the amount of traffic all the PUs generate on a single channel.

$$A = \frac{T_{ON}}{T_{ON} + T_{OFF}} \times \frac{N_p}{L} \tag{1}$$

where $T_{ON}$ and $T_{OFF}$ denote the mean ON time and OFF time of PUs, $N_p$ is the number of PUs, and $L$ is the total number of channels. In general, a higher value of the load factor (*A*) implies lower aggregate TCP throughput of the SUs.

Their simulation results showed that there is an optimum number of channels for SUs to achieve maximum aggregate throughput. The optimum number of channels depends on the number of PUs and also the number of SUs on the network. They found that the TCP in the CRN has to cope with losses caused by PU arrivals; otherwise performance is significantly deteriorated. However, their work did not address wireless link errors such as bit errors and buffer overflow. In fact, it is necessary to consider that link errors can be caused by wireless transmission impairments as well as the losses due to PU arrival, and buffer overflow on all the TCP flows of SUs.

Considering point-to-multipoint communication scheme, Khalife *et al.* [17] proposed Point-to-Multipoint Transport (PMT) protocol for single-hop CRNs. It is an acknowledgement based transport protocol which dynamically differentiates among receivers and separates them according to their reception capabilities. Further, in [18], they validated the PMT protocol over radio testbed and found that PMT can guarantee high reliability for all destinations of a group communication and adapt it flow control to the chosen channels, offering higher throughput than TCP over lossy links. However, they only considered spectrum switching, ignoring the PU activity, which is a key feature in CRNs.

### B.  Cross-layer approaches

Layering is the key design methodology in communications protocol stacks, but this strict layering may be a threat in DSA-dominated networking. This is because it should support the exchange of information which is useful for efficient communication in DSA-dominated networking. Cross-layer approaches are better than the existing strict layered approaches in achieving high network performance. This is why much existing works exploit cross layer approaches to enhance TCP performance in CRNs.

Sarkar *et al.* [19] proposed TCP Everglades (TCPE) for single-hop CRNs, which uses a cross-layer approach to serve delay-tolerant applications and to adjust the congestion window by considering spectrum sensing and bandwidth variations. The architecture of the proposed cognitive transport layer is introduced by augmenting the standard transport layer with two models: the knowledge module and cognitive module. The knowledge module is linked to the transport protocol that leverages information from the link and physical layer such as sensing times and estimated bandwidth, also the application's need. The cognitive module is for algorithms and heuristics to gather knowledge and generate control messages for managing the operation of the transport layer. They modified the congestion window adaptation in the classical TCP Westwood [20] based on the available bandwidth estimation, which is expressed

$$EBW_i = \begin{cases} EBW_{i-1} & \text{if } rtt < \beta \cdot rtt_{\min} \text{ and } OBW_i < EBW_{i-1} \\ \alpha_i \cdot EBW_{i-1} + (1 - \alpha_i) \cdot OBW_i & \text{else} \end{cases} \qquad (2)$$

where $EBW$ is the estimated available bandwidth, $OBW$ is the observed bandwidth, $\alpha$ is a static or a dynamic averaging coefficient and.

They proposed eight decision making rules for enhancing transport layer. Their approach can prevent TCP from incorrectly reducing its transmission rate during spectrum sensing. The authors assumed spectrum sensing duration is longer than RTO and extended TCP Westwood with several rules of a knowledge based module that decides TCP operation. As TCPE is designed for a single-hop cognitive wireless environment, it is not appropriate for multi-hop CRNs.

Similarly, Cheng *et al.* [21] considered the effect bandwidth variation on TCP performance in single hop CRNs. A cross-layer solution is proposed to deal with it, whose objective is to improve TCP throughput by using some useful information about the network condition from the MAC layer. They considered CR network as a centralized structure that all CR nodes communicate through the BS. CR nodes connect outside network through base BS. In addition, the cross-layer designed congestion control mechanism, Faster-Recovery is proposed. In multi-hop scenario, bandwidth variation cannot be solved by this design, for CR node cannot gather the network information far than one-hop distance.

Protocols that change lower network layers to improve throughput at the transport layers were also investigated [22-27].

Luo *et al.* [22] proposed a cross-layer design approach to jointly consider the spectrum sensing, access decision, physical-layer modulation and coding scheme, and data-link layer frame size in CRNs to maximize the TCP throughput in CRNs. The wireless channel and the primary network usage are modeled as a finite state Markov process. Due to the miss detection and the estimation error experienced by secondary users, the system state cannot be directly observed. They formulated the cross-layer TCP throughput optimization problem as a partially observable Markov decision process (POMDP). They argued that the design parameters of the CRN significantly impact TCP performance, improving it substantially if low layer parameters of the CRN are optimized jointly.

And then, the authors further improved TCP throughput through a novel multi-channel access scheme [23]. They formulated the channel access process over CRNs as a restless bandit system. With this stochastic optimization formulation, the optimal channel access policy is proposed, in which the channels with higher indices should be selected to transmit TCP traffic. However, their works did not propose any changes to the existing TCP newReno to respond to PU activities.

Considering the energy efficient, Li *et al.* [24] modified the model in [22] and presented a cross-layer scheme to improve the energy efficiency of TCP traffic considering the lower layers' characteristics, e.g., modulation, frame size, in CRNs. They use a finite state Markov channel model to characterize the fading channel and solve the optimization problem by a restless bandit approach. In addition, they gave a channel selection scheme to enhance the energy efficiency of TCP traffic.

In 2011, Wang *et al.* [25] analyzed the influence of PUs' activities and the imperfect spectrum sensing in TCP performance. This is because not only the channel unavailability, sensing overhead and sensing errors, but also the strong interaction between TCP and the lower layers will lead to the reduction of the data transmission time. It is a cross layer approach. In 2012, Wang *et al.* [26] further investigated the TCP throughput performance enhancement for CRNs through lower-layer configurations. They studied the impacts of lower-layer parameters (e.g., packet error rate, queue length, spectrum sensing accuracy), PUs activities and channel conditions on the TCP throughput. They mainly focused on the TCP throughput of SUs equipped with adaptive modulation and coding (AMC) at the physical layer and automatic repeat request (ARQ) as well as a finite size buffer at the link layer. For energy efficiency, Wang *et al.* [27] provided a model to maximize the TCP throughput and energy utility by jointly optimizing spectrum sensing time, transmit power and forward error correction (FEC) redundancy, taking the sensing errors and the interaction between TCP and lower layers into account.

However, they did not modify the congestion control mechanism in TCP Reno to respond to PU activities and spectrum sensing. Their scheme treats all types of losses as congestion losses. Hence, sometimes a TCP Reno sender unnecessarily decreases the speed at which it transmits data.

TCP-CReno [28] modifies TCP Reno so that it pauses and resumes the data when the node is performing spectrum sensing. This new protocol can obtain the current channel status by interacting with MAC layer and modify the transmission status at source. Also, the fast retransmission is improved in TCP-CReno.

Consider the DSA network as an access network to the Internet. Kumar *et al.* proposed a new network management framework, DSASync [29], to mitigate the issues that impact end-to-end connection performance when a DSA-enabled WLAN is integrated with the wired cloud. DSASync is a framework that modifies the BS's link layer that connects the outer wired network to the inner cognitive radio environment. The architecture of DSASync is shown in **Fig. 5.** DSASync sniffs the packets in transit, and maintains state information (e.g., last ACK copy, the sequence number of packets, etc.) for each ongoing TCP stream it detects. It mainly includes two components: DSASync_LL and DSASync_TCP. The former collects and maintains information about DSA parameters required by DSASync_TCP. These parameters are managed by the DSA MAC/PHY protocol, and are typically available at the link layer. DSASync_TCP is to utilize the information collected by DSASync_LL in managing both downlink and uplink TCP traffic to/from the wireless client in a DSA network.

DSASync uses information from the link layer to explicitly pause the source and destinations' TCP streams, and is focused on a 1-hop topology between the CR nodes and the BS. DSASync comprises algorithms based on buffering and traffic shaping to minimize adverse impacts on TCP/UDP connections. During DSA-related disruptions the packets are buffered in order to minimize losses. Also, the traffic rate is shaped based on the expected amount of disruptions to eliminate undesirable changes to connection behavior. There are two main advantages of DSASync-it maintains the end-to-end semantics of the standard transport protocols (TCP/UDP) and ensures compatibility by not requiring any changes to their existing implementations.

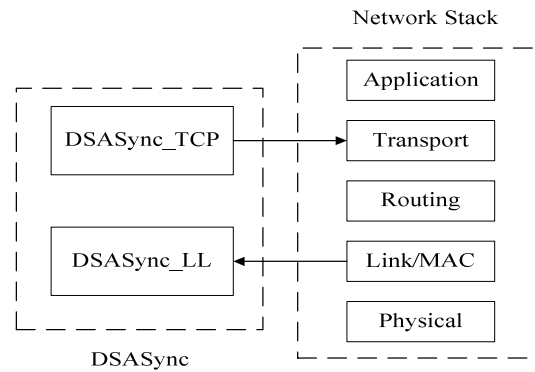

**Fig. 5.** Architectural overview of DSASync [29].

Similar to DSASync, in [30], Amjad *et al.* proposed modifications to the BS that connects TCP over the Internet to a cognitive radio network -Wireless Regional Area Network (WRAN) by modifying the BS by two proposed methods:

(a) *Local loss recovery by BS* (making BS resort to local recovery of lost frames between CRN BS and its clients) .

(b) *Split TCP connection* (modifying split TCP connection in which the BS sends crafted ACKs back to an Internet-side host on behalf of the corresponding CRN client to boost transmission speed).

They only considered the features of a CRN that could impact TCP performance: spectrum sensing time, PU activity, and the accuracy of PU's detection. The spectrum switching did not be mentioned, however, it often occurs when a PU arrives in CRNs.

In [29] and [30], the CR network is a one-hop network. The proposals restrict the modifications to the BS and are not for the transport layer in the cognitive radio nodes.

### 4.3 Transport Protocols for Multi-hop Environments

As mentioned above, their efforts have been directed towards single hop scenario. TCP performance in multi-hop CRNs is the focus of this section. It is expected that compared to one-hop CRNs, TCP will encounter more serious difficulty in providing end-to-end communications in multi-hop CRNs, such as CR ad hoc networks (CRAHNs) [31] -infrastructureless, self-organizing multi-hop networks, and lacking centralized network management. In this case, some salient characteristics of CRAHNs, which seriously deteriorate TCP performance, include the vulnerable shared media access due to opportunistic access, and the frequent route breakages due to PU arrivals, etc. For example, in sensing state, the link is in a virtual connection state; and the nodes that perform spectrum sensing cannot transmit/receive data packets. In sensing state, the source node does not know the status of intermediate nodes. Thus, it still sends data packets, and a lot of data packets have to be stored at the node preceding the sensing node. If the sensing duration is long, the cache of the node will overflow or transmitting many data packets to the next hop node after sensing state, resulting in the next hop node overflow. We term this phenomenon *flood storage effect*, which will largely reduce the TCP throughput in multi-hop CRNs.

Here, we review the state of the art in recent efforts to transport protocols for a multi-hop cognitive radio environment, for example, there are two PUs and five SUs in a CRN, as shown in **Fig. 6**.

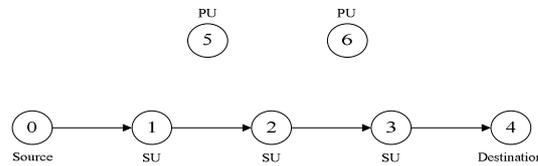

**Fig. 6.** The multi-hop network scenario.

### A. Layered approaches

Bicen *et al.* [32] explored the challenges for reliable data delivery in cognitive radio sensor networks (CRSN). They concluded that the existing transport layer protocols for wireless sensor networks and CRNs are not suitable for CRSN through extensive simulation experiments. Furthermore, open research issues for CRSN transport layer are listed. Clearly, there is a need for energy-efficient novel rate control, reliability, and congestion avoidance mechanisms which take challenges of CRSN into account. New CRSN transport protocols must consider communication impairments due to spectrum sensing and mobility such as excessive delays and packet losses incurred by cognitive cycle. As the analysis performed here clearly shows that challenges posed by DSA have direct effect on performance of transport layer, cross-layer paradigm seems promising in order to address CRSN challenges. In CRSN, its objective is conserving energy, thus, it does not optimize bandwidth utilization in [32].

In order to utilize the bandwidth more efficiently, Al-Ali *et al.* [33] proposed the first equation-based transport protocol based on TCP Friendly Rate Control for Cognitive Radio, called as TFRC-CR, which allows immediate changes in the transmission rate based on the spectrum-related changes in the network environment. TFRC-CR has the following unique features: (i) it leverages the recent FCC mandated spectrum databases with minimum querying overhead, (ii) it enables fine adjustment of the transmission rate by identifying the instances of true network congestion, as well as (iii) provides guidelines on when to re-start the source transmission after a pause due to PU activity. The TFRC-CR does not assume any cross-layer feedback or input from intermediate nodes, which aligns it with the traditional end-to-end paradigm. However, in [33], as the computation of the loss event rate is based on a history of several loss events, TFRC-CR reacts slowly to immediate decreases in loss event rate $p$. In the case of a spectrum change to an idle spectrum with larger capacity, it will therefore take a much longer time to adjust the rate to the newly available capacity, which will affect RTT and RTO values. The authors did not propose the new methods to calculate the RTT and RTO.

### B. Cross-layer approaches

In 2009, Chowdhury *et al.* [34] proposed a window-based transport protocol for CRAHNs called TP-CRAHN, which is based on the classical TCP NewReno. They modeled the transport protocol as a six-state system by using finite state machine, as shown in **Fig. 7**. They are Connection establishment, Normal, Spectrum sensing, Spectrum change, Mobility predicted, and Route failure. Each of these states addresses a particular CR network condition. The reference [34] gives the readers details about the model of TP-CRAHN.

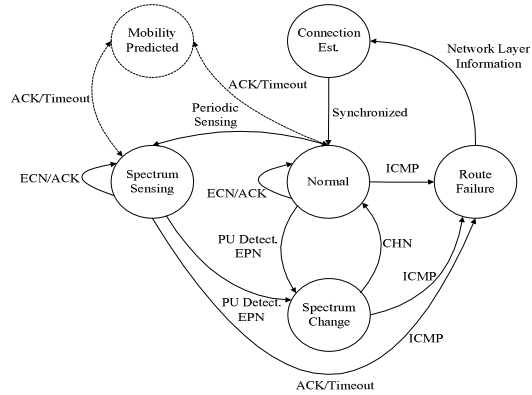

**Fig. 7.** Finite state machine model of TP-CRAHN [34].

TP-CRAHN uses additional states considering the different effects from CRAHNs, such as the delay from spectrum sensing or spectrum mobility. It also comprises additional protocol phases to adapt the protocol behavior to specific network conditions, such as sensing cycle intervals, spectrum and mobility handoffs, channel bandwidth variations, etc. In addition, TP-CRAHN exploits the Explicit Congestion Notification (ECN) mechanism to determine the loss causes. If the time lag contained in ECN is within the threshold and no prior action has been taken on an earlier ECN from the same node, TP-CRAHN assumes that congestion occurs; while any further delay indicates that the path was temporarily disconnected due to a spectrum sensing or channel switching event.

It is worth to note that in throughput analysis, the authors derived an analytical expression in TP-CRAHN, which includes three parts: normal state, spectrum sensing state, and spectrum change state [35]. To the best of our knowledge, it is the first analytical expression for throughput in CRNs, which is more suitable for CRNs with dynamic changes of channel availability depending on the PU behavior.

However, TP-CRAHN is just one modified TCP scheme which enables intermediate nodes to send or piggyback ECN to notify the source node. In addition, due to the additional states and messages, it is complex to implement and requires additional buffer space information along the path.

Koba *et al.* [36,37] first evaluated TCP CRAHN in cases where bottleneck bandwidth and RTT drastically change during communication. Based on these results, a new TCP, TCP CoBA, is proposed to further improve the throughput performance in the above cases. TCP CoBA updates the *cwnd* based upon the available buffer space in the relay node having changed its channels as well as other communication characteristics (e.g., RTT, bottleneck bandwidth).

Felice *et al.* [38, 39] extended the standard NS-2 simulator to support unique features of CRAHNs. They used the extended version of the NS-2 to evaluate various TCP versions such as TCP Reno, New Reno, Vegas and Sack in CRAHNs by considering the impact of three factors on different TCP variants: (i) spectrum sensing cycle, (ii) interference from PUs and (iii) channel heterogeneity. This is the first work that addresses the analysis of TCP over CRAHNs at both macroscopic level, i.e., computing the TCP aggregated throughput, and at microscopic level, i.e., studying the dynamics of the RTT index and congestion window (CW) size, over simulation time. However, they did not consider the impact of channel switching on TCP performance and ignore the impact of channel (re-)assignment on performance. In addition, they considered multi-flow CRAHNs, which are not mentioned in [34-37].

Considering spectrum switching frequently, which causes severe delay and high packet loss, Han *et al.* [40] first proposed a novel spectrum switching detection scheme for CRAHNs in which intermediate nodes monitor the variation of the RTT and the arrival interval time of packets to detect the spectrum switching over succeeding links. Based on the scheme, a new transport control mechanism is then designed to deal with spectrum switching in CRAHNs, which prevents source nodes from injecting more packets into the network during the spectrum switching process in order to alleviate MAC layer contentions and packet bursts.

Taking spectrum aware channel assignment and routing algorithms into account, Kim *et al.* [41] first evaluated the impact of Urban X which is a new architecture for multi-radio cognitive mesh networks on TCP. They focused on these two aspects: spectrum sensing and channel switching, which may lead to a large variation in delay and available bandwidth. They varied different characteristics such as the load and intensity of external traffic and found optimum parameters for spectrum sensing and channel switching configuration. And then, they proposed a new transport control protocol TP-UrbanX [42] for the Urban-X, which uses information available from back-pressure scheduling to detect network congestion and TCP ACKs considering the intrinsic properties of the environment such as channel switching, spectrum sensing and external interference. It integrates reinforcement based learning techniques in order to adapt its operation over time to changing network conditions.

Singh *et al.* [43] proposed a cross-layer distributed mechanism using link layer triggers from cognitive radio modules and Freeze TCP [44] to optimize the performance of TCP over CRNs for the train. The proposed mechanism uses a cross-layer distributed approach in which a connection manager is installed on clients wanting to use Internet connectivity on trains. The connection manager manages and optimizes the TCP connection without changing TCP stack and without breaking compatibility with standard TCP implementations. The performance of the proposed mechanism was evaluated and it showed significant improvement in terms of link utilization efficiency as compared to standard TCP. However, they did not analyze the effect of PU activities on TCP performance, which is a key factor in improvement TCP performance over CRNs.

All of above mentioned transport protocols for CRNs have assumed that either all nodes are in CRNs [32-42] or the TCP sender side is in CR links [7, 16, 19, 21-30] or the TCP receiver side is in CR links [15, 43]. In these proposals, lower layer information such as spectrum sensing status could be easily exploited to adjust the congestion window for maximizing the throughput. However, when the TCP sender is located remotely and connected by a wire-line over the Internet, the access network to the TCP receiver is based on a CR link, the lower layer information cannot be exploited. Under this scenario, Yang *et al.* [45] proposed an enhanced TCP for CR networks called TCP-CR to improve the existing TCP by (1) detection of primary user (PU) interference by a remote sender without support from lower layers, (2) delayed congestion control (DCC) based on PU detection when the RTO expires, and (3) exploitation of two separate scales of the congestion window adapted for PU activity. With the development of access network technology, this topology will play a key role in our daily lives. This scenario is a more realistic deployment scenario.

TCP relies on end-to-end retransmission to provide reliability for transmission, which basically consumes more bandwidth and needs longer time than hop-by-hop retransmission.

The above transport protocols did not consider the method to deal with the loss in data transmission. Network coding [46] can mask losses from the congestion control over lossy networks [47]. Based on this, in [48], we investigate the limitations of TCP in multi-channel multi-radio multi-hop CRNs, and propose a novel transmission control protocol TCPJGNC (TCP Joint Generation Network Coding, JGNC [49]-our previous work), a network coding-based TCP protocol.

In JGNC, suppose the data is divided into $m$ generations, and each generation has $k$ packets and be independently encoded. A coded packet $c_i$ is a random linear combination of the original packets. That is $c_i = \sum_{h=1}^{k} e_{ih} P_h$, where the coefficient $e_{ih}$ is a randomly chosen element in a finite field $F$ [50].

Clearly, the entire data block has n = $m \times k$ packets. The destination only needs to receive $k$ independent packets for decoding the original packets over each generation. When the destination receives less than $k$ independent packets belonging to the first generation, the source will not transmit first generation again; instead, it combines this generation with the second generation to be encoded, then transmit. When the destination receives $2k$ independent packets, it can decode these two generation packets. If not, all the packets of these two generations are encoded with the packets of the third generation; when the destination receives $3k$ independent packets, it can decode the packets. And so on, until $k$ generation packets are encoded. Of course, if the destination could receive enough packets to decode a generation, then it is not necessary to joint this generation with next generation to be encoded.

In [48], we dynamically adjust the number of packets involved in the network coding operations according to the wireless communication environment in TCPJGNC. In the meantime, we also provide a scheme to reduce the retransmission by adopting JGNC in TCPJGNC. To the best of our knowledge, TCPJGNC is the first transmission control protocol for CRNs from a network coding perspective. An analysis of approximate expected throughput in TCPJGNC is provided and simulation results indicate that TCPJGNC can significantly improve the network performance in terms of throughput and bandwidth efficiency.

Considering without a common control channel (CCC) in CRAHNs, Song *et al.* [51] proposed a comprehensive end-to-end congestion control framework in CRAHNs considering the non-uniform channel availability. They studied two types of existing congestion control methods: explicit feedback mechanism without timeouts and the timeout mechanism. Through extensive simulations, they found that none of the existing transport layer approaches can be simply applied to effectively address the congestion issue in CRAHNs. However, they did not give a method for solving congestion issue in CRAHNs, also, their framework mainly relied on extensive feedback, which is more difficult to obtain the feedback information in large-scale CRAHNs.

Based on single hop and multi-hop environments, **Table 1** provided in this section summarizes the different proposals in more detail for quick overview. This tabular representation specifies the different characteristics of each proposal. **Table 1** shows that the classification makes it easy to compare approaches that fall under the same category. However, it is difficult to present a comprehensive comparison of all the approaches together because each one addresses specific problems.

**Table 1**. Comparison of transport protocols in CRNs.

| | Proposals | Characteristic | Flood storage effect | Compatibility with the Internet | Layered | Cross-layer | TCP mechanism | Lower parameters | PU activity | Bandwidth variation | Spectrum switching | Energy | Loss recovery |
|---|---|---|---|---|---|---|---|---|---|---|---|---|---|
| single hop cognitive radio environments | Issariyakul *et al.* [5] | Consider a new type of loss called service interruption, loss for TCP NewReno over CRNs. | - | - | Y | - | - | - | Y | Y | - | - | - |
| | Slingerland *et al.* [15] | Evaluate TCP (SACK, NewReno, Vegas) performance in DSA scenarios varying link capacity and PU detection errors. | - | - | Y | - | - | - | Y | Y | - | - | - |
| | Kondareddy *et al.* [16] | Modify the analytical model proposed in [15] to incorporate the delay caused by PU's and SU's traffic and PU detection errors. | - | - | Y | - | - | - | Y | - | - | - | - |
| | PMT [17,18] | An acknowledgement based transport protocol which dynamically differentiates among receivers and separates them according to their reception capabilities. | - | - | Y | - | - | - | - | Y | Y | - | - |
| | TCPE [19] | Uses a cross-layer approach to serve delay-tolerant applications and to adjust the congestion window by considering spectrum sensing and bandwidth variations. | - | - | - | Y | Y | Y | - | Y | Y | - | - |
| | Cheng *et al.* [21] | A cross-layer solution considering the effect bandwidth variation on TCP performance in single hop CRNs. | - | - | - | Y | Y | Y | - | Y | - | - | - |
| | Luo *et al.* [22,23] | A cross-layer design approach to jointly consider the spectrum sensing, access decision, physical-layer modulation and coding scheme, and data-link layer frame size to maximize the TCP throughput in CRNs. | - | - | - | Y | - | Y | - | - | - | - | - |
| | Li *et al.* [24] | A cross-layer scheme to improve the energy efficiency of TCP traffic considering the lower layers' characteristics, e.g., modulation, frame size. | - | - | - | Y | - | Y | - | - | - | Y | - |
| | Wang *et al.* [25,26] | Investigate the TCP throughput performance enhancement for CRNs considering PUs' activity and lower-layer configurations. e.g., packet error rate, queue length. | - | - | - | Y | - | Y | Y | - | Y | - | - |
| | Wang *et al.* [27] | Maximize the TCP throughput and energy utility by jointly optimizing spectrum sensing time, transmit power and FEC redundancy. | - | - | - | Y | - | Y | Y | - | Y | Y | - |
| | TCP-CReno [28] | It pauses and resumes the data when the node is performing spectrum sensing. It can obtain the current channel status by interacting with MAC layer and modifies the transmission status at source. Also, the fast retransmission is improved. | - | - | - | Y | Y | Y | - | - | Y | - | - |
| | DSASync [29] | Mitigate the issues that impact end-to-end connection performance when a DSA-enabled WLAN is integrated with the wired cloud. | - | Y | - | Y | Y | Y | - | Y | Y | - | - |
| | Amjad *et al.* [30] | Modify the BS that connects TCP over the Internet to a CRN by Local loss recovery by BS and Split TCP connection. | - | Y | - | Y | - | Y | - | - | - | - | Y |
| multi-hop cognitive radio environments | Bicen *et al.* [32] | Conclude that the new CRSN transport protocols must be considered energy-efficient novel rate control, reliability, and congestion mechanisms. | Y | - | Y | - | - | Y | Y | Y | Y | Y | - |
| | TFRC-CR [33] | An equation-based transport protocol based on TCP Friendly Rate Control for Cognitive Radio, which allows immediate changes in the transmission rate based on the spectrum-related changes in the network environment. | Y | - | Y | - | Y | - | Y | - | - | - | - |
| | TP-CRAHN [34,35] | A window-based transport protocol for CRAHNs, modelling the transport protocol as a six-state system by using finite state machine, Connection establishment, Normal, Spectrum sensing, Spectrum change, Mobility predicted, and Route failure. | Y | - | - | Y | - | Y | Y | Y | Y | - | - |
| | TCP CoBA [36,37] | A new TCP for CRAHNs where bottleneck bandwidth and RTT drastically change during communication. | Y | - | - | Y | - | Y | - | - | Y | - | - |
| | Felice *et al.* [38,39] | Evaluate various TCP versions in CRAHNs by considering the impact of three factors on different TCP variants: spectrum sensing cycle, interference from PUs and channel heterogeneity. | Y | - | - | Y | - | Y | Y | - | Y | - | - |
| | Han *et al.* [40] | A novel spectrum switching detection scheme for CRAHNs in which intermediate nodes monitor the variation of the RTT and the arrival interval time of packets to detect the spectrum switching over succeeding links. | Y | - | - | Y | - | Y | - | - | Y | - | - |
| | TP-UrbanX [41,42] | Use information available from back-pressure scheduling to detect network congestion and TCP ACKs considering the intrinsic properties of the environment. It integrates reinforcement based learning techniques in order to adapt its operation over time to changing network conditions. | Y | - | - | Y | - | Y | - | Y | Y | - | - |
| | Singh *et al.* [43] | A cross-layer distributed mechanism using link layer triggers from cognitive radio modules and Freeze TCP to optimize the performance of TCP over CRNs for the train. | Y | Y | - | Y | - | Y | - | - | - | - | - |
| | TCP-CR [45] | An enhanced TCP for CRNs using by (1) detection of PU interference by a remote sender without support from lower layers, (2) delayed congestion control based on PU detection when the RTO expires, and (3) exploitation of two separate scales of the congestion window adapted for PU activity. | Y | Y | - | Y | - | Y | - | Y | - | - | - |
| | TCPJGNC [48] | It is the first transmission control protocol for CRNs from a network coding perspective, dynamically adjusting the number of packets involved in the network coding operations according to the wireless communication environment. | Y | - | - | Y | Y | Y | Y | Y | Y | - | Y |
| | Song *et al.* [51] | A comprehensive end-to-end congestion control framework in CRAHNs without CCC considering the non-uniform channel availability. | Y | - | - | Y | Y | Y | - | Y | - | - | - |

(The '-' mark is used if the approach does not involve a particular aspect, otherwise 'Y' is used.)

## 5. Research Challenges and Open Issues

In previous sections, we have reviewed the state-of-the-art in the transport protocols of CRNs. In this section we discuss some of open issues (but not limited to) for which searching for a better solution demands special efforts.

- Compatibility with the Internet. For the purpose of internetworking with the wired Internet as required in future CRNs based pervasive computing; the TCP designed for CRNs should be fully compatible with the Internet. In [29, 30, 43, 45], the authors considered this scenario, however, some issues such as retransmission mechanism need to be studied further.

- Channel uncertainty requires novel congestion avoidance algorithms. More PU appearances may increase the probability of congestion due to service interruption loss. Thus, congestion control mechanism should take spectrum sensing duration, spectrum assignment and dynamically changing PU activity into account. Hence, cross-layer optimizations considering spectrum management and congestion control mechanism are highly desirable.

- Most of the existing proposals either address congestion control or optimize lower-layer parameters. Loss recovery is rarely considered in the above works, none of them except [30, 48] investigates both congestion control and loss recovery systematically. In fact, loss recovery no doubt can enhance end-to-end reliability while a proper congestion control can reduce packet loss and provide better throughput. Therefore, when designing transport protocols for CRNs, it should take congestion control, lower-layer parameters and loss recovery into account for performance optimization on the spectrum-efficiency and other performance metrics.

- Existing schemes only consider single path routing. When multi-path routing/opportunistic routing [52] is used in the network layer, packets coming from different paths may not arrive at the receiver in order. TCP receiver may misinterpret such out-of-order packet arrivals as a congestion signal. Hence, the receiver will send duplicate ACKs that inform the sender to invoke congestion control algorithms. The sender reduces its transmission rate unnecessarily, leading to low throughput. Future work is needed to ensure that the characteristics of paths (e.g., channel availability, bandwidth asymmetry) to the receiver or to multiple receivers are taken into account. In addition, fairness should be studied further, which should consider channel availability and bandwidth efficiency.

- Bandwidth variation requires new window-based mechanisms for TCP congestion control. In CRNs, bandwidth oscillation readily occurs in spectrum changing. In order to provide better throughput, TCP should employ a new dynamic window adjusting scheme to adapt effectively to bandwidth variations. As mentioned above, some existing works considered bandwidth variation, however, they only updated window size according bandwidth, which ignored PU activity and SU activity. How to design novel window-based mechanisms for TCP congestion control according to the features of CRNs is an open issue.

## 6. Conclusion

In this paper, the transport protocols for CRNs have been summarized. According to the unique features of CRNs, we analyze the challenges in designing transport protocols over CRNs. We classify the existing transport protocols into two categories based on transmission environments (single hop or multi-hop), and in each category, the existing proposals are described and analyzed from two approaches, layered and cross-layer. Although some transport protocols have been proposed, there are some possible future works need further study. We hope this survey will provide a better understanding of the literature of transport protocols in CRNs and spark new research interests and developments in this area.

Zhong et al.: Transport Protocols in Cognitive Radio Networks: A Survey

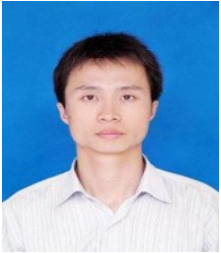

**Xiaoxiong Zhong** received his M.S degree in Computer Science at Lanzhou University of Technology, Lanzhou, China, in 2010. He is currently working toward a Ph.D degree with the Department of Computer Science and Technology, Harbin Institute of Technology, Shenzhen Graduate School, China. His research interests include protocol analysis and resource management in cognitive radio networks.

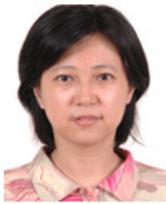

**Yang Qin** received her B.Sc (with first class honors) in Computer Science at Southwest Jiaotong University (China), in 1989, M.S in Computer Science at Huazhong University of Science & Technology, Wuhan, Hubei, in 1992, and Ph.D in Computer Science, Hong Kong University of Science & Technology, Kowloon, Hong Kong at November of 1999. From 1999 to 2000, she has visited the Washington State University as a Postdoc, USA. From 2000 to 2008, she is an assistant professor Nanyang Technological University, Singapore. Currently, she is an associate professor in the Department of Computer Science and Technology, Harbin Institute of Technology, Shenzhen Graduate School, China. Her research interest is in the area of wireless networks, mobile computing, cross-layer design, QoS of routing and scheduling, high speed optical networks and so forth. She is a senior member of IEEE.

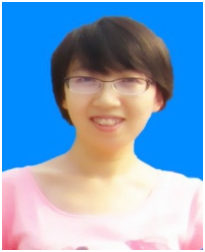

**Li Li** received her B.S and M.S degrees in Computer Science at Lanzhou University of Technology, Lanzhou, China, in 2007 and 2011, respectively. She is currently working toward a Ph.D degree with the Department of Computer Science and Technology, Harbin Institute of Technology, Shenzhen Graduate School, China. Her research interest is in the area of wireless networks, cross-layer design, and opportunistic networks.